\documentclass{WileyMSP-template}
\usepackage{graphicx, color, hyperref, amsfonts, amsmath, bm, multirow}
\usepackage{lastpage}

\usepackage{cite} 

\newcommand{\figref}[1]{Figure~\ref{#1}}
\newcommand{\tabref}[1]{Table~\ref{#1}}

\newcommand{\secref}[1]{Section~\ref{#1}}
\newcommand{\Eqref}[1]{Equation~(\ref{#1})}

\newcommand{\argmax}{\mathop{\rm arg~max}\limits}

\newcommand{\x}{\bm{x}}
\newcommand{\z}{\bm{z}}
\newcommand{\y}{\bm{y}}
\newcommand{\lnp}[2]{\mathrm{ln}P_{\bm{#1}}(#2)}
\newcommand{\px}{P_{\bm{\theta}}(\x)}
\newcommand{\pz}{P_{\bm{\theta}}(\z)}

\newcommand{\pzlx}{P_{\bm{\theta}}(\z|\x)}
\newcommand{\pxlz}{P_{\bm{\theta}}(\x|\z)}
\newcommand{\pt}[1]{P_{\bm{\theta}}({#1})}

\newcommand{\qzlx}{Q_{\bm{\phi}}(\z|\x)}
\newcommand{\intz}[1]{\int {#1} d\z}
\newcommand{\vlb}{L_{\bm{\theta}, \bm{\phi}}(\x)}
\newif\ifDraft
\Drafttrue

\definecolor{DarkGreen}{rgb}{0.0,0.45,0}
\definecolor{Sakujo}{rgb}{0.9,0.5,0.9}
\definecolor{Grey}{rgb}{0.7,0.7,0.7}
\ifDraft

\else

\fi

\begin{document}
\pagestyle{fancy}
\rhead{\includegraphics[width=2.5cm]{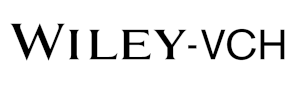}}
\chead{\thepage{}/{}\pageref{LastPage}}

\title{Training Process of Unsupervised Learning Architecture \\
for Gravity Spy Dataset}
\date{}

\author[1,+]{Yusuke~Sakai}
\author[2, 3]{Yousuke~Itoh}
\author[4]{Piljong~Jung}
\author[5]{Keiko~Kokeyama}
\author[6]{Chihiro~Kozakai}
\author[7]{Katsuko~T.~Nakahira}
\author[8]{Shoichi~Oshino}
\author[9, 10, 11]{Yutaka~Shikano}
\author[1, 12, 13, ++]{Hirotaka~Takahashi}
\author[8]{Takashi~Uchiyama}
\author[7]{Gen~Ueshima}
\author[6]{Tatsuki~Washimi}
\author[8]{Takahiro~Yamamoto}
\author[8]{Takaaki~Yokozawa}

\affil[1]{Research Center for Space Science, Advanced Research Laboratories, Tokyo City University, Setagaya-ku, Tokyo 158-0082, Japan}
\affil[2]{Graduate School of Science, Osaka Metropolitan University, Sumiyoshi-ku, Osaka City, Osaka 558-8585, Japan}
\affil[3]{Nambu Yoichiro Institute of Theoretical and Experimental Physics (NITEP),
Osaka Metropolitan University, Sumiyoshi-ku, Osaka City, Osaka 558-8585, Japan}
\affil[4]{National Institute for Mathematical Sciences, Daejeon 34047, Republic of Korea}
\affil[5]{School of Physics and Astronomy, Cardiff University, The Parade, Cardiff, CF24 3AA United Kingdom}
\affil[6]{Gravitational Wave Science Project, Kamioka branch, National Astronomical Observatory of Japan, Hida City, Gifu 506-1205, Japan}
\affil[7]{Department of Information and Management Systems Engineering, Nagaoka University of Technology, Nagaoka, Niigata 940-2188, Japan}
\affil[8]{Institute for Cosmic Ray Research, KAGRA Observatory, The University of Tokyo, Hida City, Gifu 506-1205, Japan}
\affil[9]{Graduate School of Science and Technology, Gunma University, Maebashi, Gunma 371-8510, Japan }
\affil[10]{Institute for Quantum Studies, Chapman University, Orange, CA 92866, USA}
\affil[11]{JST PRESTO, Kawaguchi, Saitama 332-0012, Japan}
\affil[12]{Institute for Cosmic Ray Research, The University of Tokyo, Kashiwa City, Chiba 277-8582, Japan}
\affil[13]{Earthquake Research Institute, The University of Tokyo, Bunkyo-ku, Tokyo 113-0032, Japan}

\affil[+]{g2191402@tcu.ac.jp}
\affil[++]{hirotaka@tcu.ac.jp}

\maketitle



\begin{snugshade}
  \begin{abstract}
Transient noise appearing in the data from gravitational-wave detectors frequently causes problems, such as instability of the detectors and overlapping or mimicking gravitational-wave signals. Because transient noise is considered to be associated with the environment and instrument, its classification would help to understand its origin and improve the detector’s performance. In a previous study, an architecture for classifying transient noise using a time–frequency 2D image (spectrogram) is proposed, which uses unsupervised deep learning combined with variational autoencoder and invariant information clustering. The proposed unsupervised-learning architecture is applied to the Gravity Spy dataset, which consists of Advanced Laser Interferometer Gravitational-Wave Observatory (Advanced LIGO) transient noises with their associated metadata to discuss the potential for online or offline data analysis. In this study, focused on the Gravity Spy dataset, the training process of unsupervised-learning architecture of the previous study is examined and reported.
  \end{abstract}
\end{snugshade}
  


\section{Introduction} \label{sec:introduction}
Advanced Laser Interferometer Gravitational-Wave Observatory (Advanced LIGO) detector~\cite{ref:aasi2015advanced} located at Livingston and Hanford, USA, made its first observation of gravitational waves from the coalescence of a binary black hole in September 2015~\cite{ref:Abbott2016}.
Following that, Advanced LIGO and Advanced Virgo~\cite{ref:acernese2014advanced} in Pisa (Italy) have made three international joint observations and observed as many as 90 events of gravitational waves emitted by the coalescence of the compact binary~\cite{ref:Abbott2019,ref:Abbott2020,ref:Abbott2021_a,ref:Abbott2021_b}.
KAGRA~\cite{ref:akutsu2018kagra,ref:2021PTEP.2021eA101A,ref:2020arXiv200802921K,ref:2021PTEP.2021eA102A,ref:2022bkagra,ref:2022akagra} in Japan will join the next (fourth) observing run (O4) by Advanced LIGO and Advanced Virgo.

In the data analysis of gravitational waves, the technique of separating the gravitational waves from the noise in the observed data is essential because the signals of the gravitational waves are generally smaller than the detector noise. Because the gravitational-wave detector is sensitive to environmental and instrumental conditions, such as ground motions, air pressure, optics suspensions, laser fluctuations, vacuum, and mirror, the non-stationary and non-Gaussian noise called ``transient noise'' frequently appears in the detector's data~\cite{ref:Nguyen_2021}.
The transient noise causes the detector to be unstable, it can also hide and imitate gravitational-wave signals~\cite{ref:Davis_2021}.
During O3b~\cite{ref:Abbott2021_b}, LIGO Scientific Collaboration, Virgo Collaboration, and KAGRA Collaboration (LVK) reported that the transient noise rate with a signal-to-noise ratio (SNR) of $> 6.5$ was $1.17$ events per minute at LIGO Livingston.

Machine learning is increasingly being applied in the study of transient noise~\cite{ref:Biswas_2013,ref:Powell_2015,ref:Mukund_2017}.
Transient noise has various time-frequency characteristics that are related to its causes. Classifying transient noise could provide a clue to explore its origins and improve the performance of the detector~\cite{ref:Soni_LSC_2021}. Thus, the Gravity Spy project~\cite{ref:Zevin2017,ref:Bahaadini2018,ref:Soni2021,ref:Bahaadini2017} attempted to classify the transient noise. In the Gravity Spy project, the Omicron software~\cite{ref:Robinet2020} was used to identify the signal of transient noise observed in the time-series data. Following that, using the Omega Scan~\cite{ref:Chatterji2004}, a time-frequency spectrogram was created around the identified transient noise as a 2D image. Based on a portion of these created 2D images, 22 types of labels associated with the characteristics or causes of transient noise were annotated for the analysis using cloud resources in collaboration with LIGO detector characterization experts and volunteer citizen scientists. Both the images and labels were recorded. Finally, using the pre-classified images and labels, they classified the transient noise in the remaining images using supervised learning. 

Recently, there are some reports on the clustering and/or classification of transient noise using unsupervised learning ~\cite{ref:sakai2021unsupervised,ref:george2018,ref:bahaadini2018direct,ref:Bini2020unsupervised,ref:Ramezani_2021}.
Sakai et al.~\cite{ref:sakai2021unsupervised} discussed an architecture for the classification of transient noise using unsupervised learning, which combines a variational autoencoder (VAE)~\cite{ref:Kingma2014}~\cite{ref:Kingma2019} and invariant information clustering (IIC)~\cite{ref:Ji2019}.
The consistency between the label annotated by the Gravity Spy project and the class provided by the proposed unsupervised-learning architecture was confirmed using the Gravity Spy dataset of LIGO O1, and the potential for the classification of transient noise using unsupervised learning was discussed. Unsupervised learning is expected to reduce annotation work for training data, increase classification objectivity, and even classify a new class, such as the transient noise because it does not require any pre-assigned labels for the training dataset.
Moreover, the training process of unsupervised learning is essential and some detailed reports on the training process can be found in refs.~\cite{ref:Choi_2019,Wilson_2017}.
In this study, focused on the Gravity Spy dataset, the training process of unsupervised-learning architecture of ref.~\cite{ref:sakai2021unsupervised} is examined and reported. 

The remainder of this study is organized as follows:
In Section~\ref{sec:method}, we explain the outline of VAE and IIC, which are used in the proposed architecture.
In Section~\ref{sec:architecture}, we review the architecture, which was proposed in our previous research~\cite{ref:sakai2021unsupervised}.
In Section~\ref{sec:training}, we report on how the training process of the unsupervised learning architecture was conducted.
We also give a summary of the evaluation and obtained results discussed in~\cite{ref:sakai2021unsupervised}.
Section~\ref{sec:summary} presents a summary of the work.

\section{Outline of Used Method for Proposed Architecture} \label{sec:method}
The proposed architecture discussed in Section~\ref{sec:architecture} to classify transient noise is combined VAE~\cite{ref:Kingma2014}~\cite{ref:Kingma2019} and IIC~\cite{ref:Ji2019}, both of which are known as unsupervised-learning methods.

\subsection{Variational Autoencoder}
We introduce VAE~\cite{ref:Kingma2014}~\cite{ref:Kingma2019} that forms the feature learning~\cite{ref:Zhong2016}~\cite{ref:Bengio2013} part of our proposed architecture.
VAE is a Bayesian inference method that uses a predictive distribution for the parameters and was designed for unsupervised learning.
VAE has an architecture that compresses an image to the latent variables before reconstructing the original image from the latent variables.
Let $\mathcal{X} \subset \mathbb{R}^D$ be the input space and $\mathcal{Z} \subset \mathbb{R}^J$ be the latent space, where $D, J \in \mathbb{Z}$ and $J < D$.
Suppose a true (but unknown) probability distribution is the parameterized model $P_{\bm{\theta}}$ of the variable $\x \in \mathcal{X}$, and the latent variable $\bm{z} \in \mathcal{Z}$ is represented instead of the parameters $\bm{\theta}$, which are not directly observed.
The marginal likelihood $\px$, which we are interested in, has the following relation (Bayes' theorem), $\px = \pxlz \pz/\pzlx$.
Then, let the likelihood $\pxlz$ be the generative model, and a prior distribution $\pz$ be hypothesized to be a Gaussian distribution.
Unfortunately, $\px$ is \textit{intractable}; therefore, we cannot evaluate it directly.
However, considering the posterior distribution as the inference model $\pzlx \sim \qzlx$ that is parameterized by $\bm{\phi}$, we can evaluate $\px$ indirectly.
Then, the logarithm marginal likelihood can be expressed as
\begin{equation}
  \lnp{\bm{\theta}}{X}  = D_{\mathrm{KL}} \left(\qzlx || \pzlx \right) + \vlb
  \label{eq:logarithm-marginal-likelihood},
\end{equation}
where $D_{\mathrm{KL}}$ is the Kullback-Leibler divergence of two distributions, and $\vlb$ is known as the variational lower bound, respectively. The first term on the right-hand side of \Eqref{eq:logarithm-marginal-likelihood} gradually approaches zero as the accuracy is increased.
Therefore, maximizing the log-likelihood can be replaced by the problem of maximizing the variational lower bound as follows,
$\argmax_{\bm{\theta}} \lnp{\bm{\theta}}{\x}=\argmax_{\bm{\theta}, \bm{\phi}} \vlb$.
Then, the equation for the variational lower bound is expressed as
\begin{eqnarray}
  \vlb \sim -D_{\mathrm{KL}}\left(
  \qzlx || \pt{\z}
  \right)
  + \intz{\qzlx \ln \pxlz},
  \label{eq:variational-lower-bound}
\end{eqnarray}
where $\mathcal{X} = \{ \x = f_{\bm{\theta}}(\z) | \z \in \mathcal{Z}\}$ represents the input space and a map $f_{\bm{\theta}}:\mathcal{Z} \to \mathcal{X}$ is known as a decoder.
Similarly, the latent space is also represented by $\mathcal{Z} = \{ \z = g_{\bm{\phi}}(\x) | \x \in \mathcal{X}\}$ and a map $g_{\bm{\phi}}:\mathcal{X} \to \mathcal{Z}$ is is known as an encoder.
Suppose the two multivariate Gaussian distributions in the latent space are expressed as follows,
$\qzlx = N(\bm{z};\bm{\mu}, \bm{\sigma}^2\bm{I})$ and $\pz = N(\bm{z}; \bm{0}, \bm{I})$, where $\bm{\mu}$ is the mean and $\bm{\sigma}$ is the variance, and $\bm{I}$ is the identity matrix.
Then, using \textit{the reparameterization trick}~\cite{ref:Kingma2014}, \Eqref{eq:variational-lower-bound} can be expressed as
\begin{eqnarray}
  \vlb =
  \frac{1}{2}\sum_{j=1}^J ( 1 + \ln \sigma_j^2 - \mu_j^2 - \sigma_j^2)
  + \frac{1}{L} \sum_{l=1}^{L} \lnp{\bm{\theta}}{\x| \z^{(l)}},
  \label{eq:variational-lower-bound-formula}
\end{eqnarray}
where $\mu_j$ and $\sigma_j$ are the mean and variance of $J$-dimensional Gaussian distribution, respectively, and $L$ is the number of samples per datapoint.
$L$ can be set to $1$ when the minibatch size is large enough~\cite{ref:Kingma2014}.

\subsection{Invariant Information Clustering} \label{sec:iic}
We briefly explain IIC~\cite{ref:Ji2019} that forms the classification part of our proposed architecture.
Although classification using supervised learning generally requires numerous labels for its training, IIC can classify without these labels and achieve results comparable to supervised learning.
The goal of IIC is to learn what is in common between the paired data.
These paired data, for example, could be different images with the same characteristics.
Let $\x, \x^{\prime} \in \mathcal{X}$ be paired data following a joint probability distribution $P(\x, \x^{\prime}$) and $C \in \mathbb{Z}$ be the number of output classes.
IIC learns a classifier (representation) $\bm\Phi:\mathcal{X} \to \mathbb{R}^C$ that maximizes the mutual information $I$ expressed by entropy $H$ as follows
\begin{equation}
  I(\bm\Phi(\x), \bm\Phi(\x^{\prime})) = H(\Phi(\x)) - H (\Phi(\x)|\Phi(\x^{\prime})).
  \label{eq:mutual_information}
\end{equation}
Moreover, the maximum value of \Eqref{eq:mutual_information} such that $\x = \x^{\prime}$ can be analytically obtained as follows
\begin{equation}
  \max(I(\bm\Phi(\x)) = \ln C.
  \label{eq:maximum_mutual_information}
\end{equation}

Calculating a criterion for data within the same class and data between different classes is useful for evaluating the performance of the classifier.
Thus, let $P_{ij} = \bm\Phi(\x^{(i)}) \cdot \bm\Phi(\x^{\prime(j)})^T$ be the conditional joint distribution and \\ 
$P_i = \sum_j \bm\Phi(\x^{(i)}) \cdot \bm\Phi(\x^{\prime(j)})^T$ be the marginal distribution, where $\x^{(i)} \in \mathcal{X}$ are data belonging to the $i$th class, and a notation $T$ means a transpose and ``$\cdot$'' means an inner product.
Then, the mutual information \Eqref{eq:mutual_information} can be expressed as
\begin{equation}
  I(\bm\Phi(\x), \bm\Phi(\x^\prime))
  = \sum_i^C\sum_j^C P_{ij}\ln \frac{P_{ij}}{P_i P_j}.
  \label{eq:ojective_mutual_information}
\end{equation}
The objective of training IIC is to construct the classifier $\Phi$ that maximizes \Eqref{eq:ojective_mutual_information}.

\section{Outline of Proposed Architecture} \label{sec:architecture}
\begin{figure}[t]
  \centering
  \includegraphics[scale=0.8]{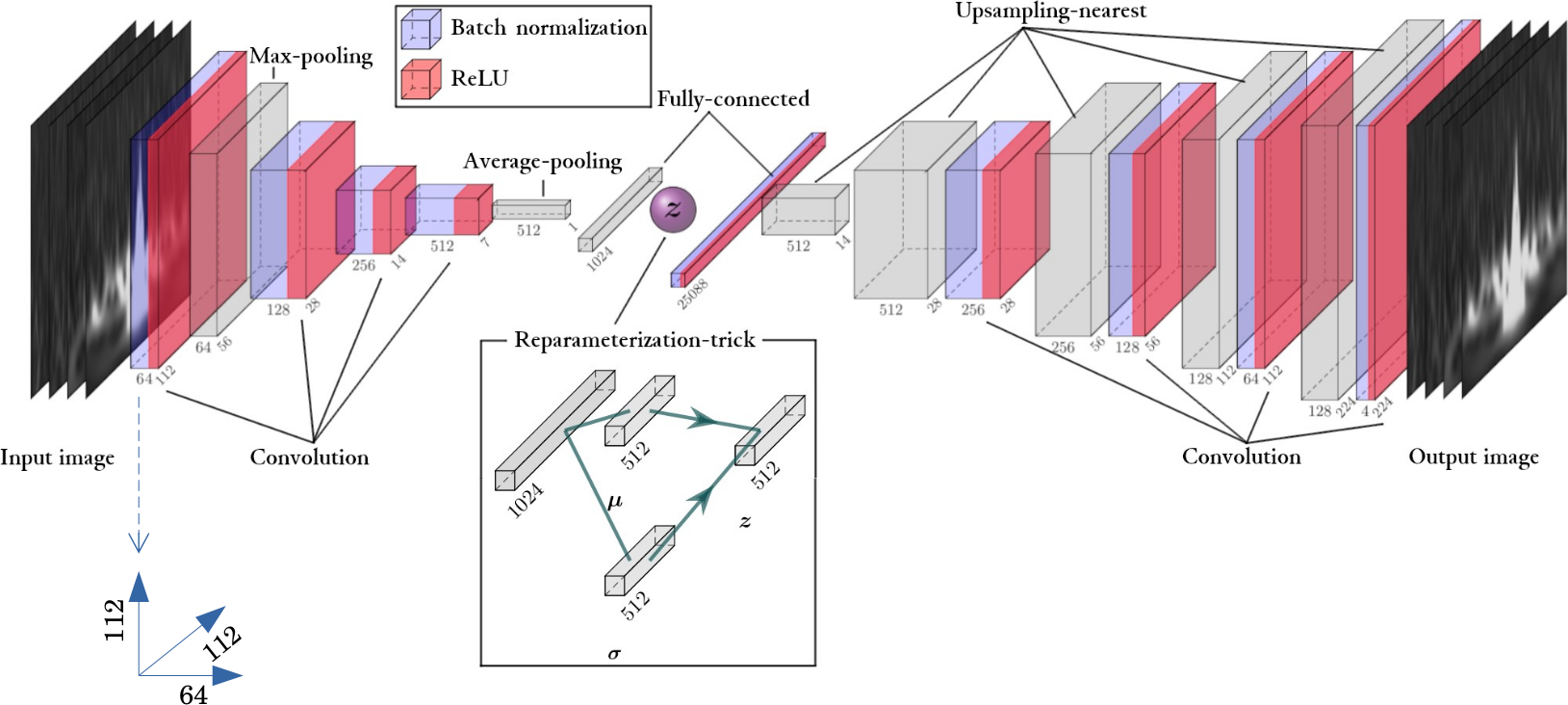}
  \caption{
    Schematic view of the proposed VAE architecture for training the latent variables from input images of transient noise.
    The encoder part is from the input layer (at the left end) to the latent variables $\z$ (at the center), and the decoder part is from $\z$ to the output layer (at the right end).
    The blue block represents batch normalization, whereas the red block represents the ReLU activation function.
    A solid line in reparameterization-trick blocks simply means splitting the component in half, and lines with an arrow mean the sampling from the mean $\bm\mu$ and the variance $\bm\sigma$.
  }\label{fig:vae_arch}
\end{figure}

We review an unsupervised-learning architecture proposed in ref.~\cite{ref:sakai2021unsupervised}, which has deep convolutional neural networks for the classification of transient noise.
The proposed architecture consists of two processes: feature learning using VAE and classification using IIC.
We input four 2D images of transient noise with different time durations to the proposed architecture, and its shape is $(4, 224, 224)$ data having four square images $(224, 224)$.
More details of the input 2D images are shown in \secref{sec:outline-of-target-dataset}.
Regarding the implementation of VAE, we used the architecture, as shown in \textbf{\figref{fig:vae_arch}}.
Each block shows the components of neural networks and their shape.
The purpose of VAE is to learn the latent variables, which are the compressed features from a large dataset.
We used four convolutional neural networks in our proposed architecture because convolutional layers are known to be effective in image processing.
Additionally, we used a ReLU activation function to avoid the vanishing gradient problem in deep neural networks and used a batch normalization technique in all convolutional layers to stabilize training.
For the encoder part, the first convolution layer outputs a shape of $(M, 64, 112, 112)$ from the input 2D image having a shape of $(M, 4, 224, 224)$, where $M$ is a minibatch size.
A max-pooling layer next to the convolutional layer extracts the features and compresses them into a shape of $(M, 112, 56, 56)$.
Following that, further features are extracted with three convolutional layers and are averaged through an average-pooling layer.
A fully connected layer connects to all variables in neural networks.
After using \textit{reparameterization trick}~\cite{ref:Kingma2014}, the latent variables $\z$ are obtained (in the case of \figref{fig:vae_arch}, its shape is $(512)$).
For the decoder part, the decoder constructs an image element whose shape is $(M, 512, 14, 14)$ from the latent variables $\z$ through a fully connected layer and an upsampling nearest layer.
Because 2D nearest-neighbor upsampling simply doubles the shape of the input, this layer is combined with a convolutional layer to generate an image.
The decoder has four upsampling and convolutional layers.
Following that, a $(M, 4, 224, 224)$ image with the same shape as the input was obtained.

\begin{figure}[t]
  \centering
  \includegraphics[scale=0.7]{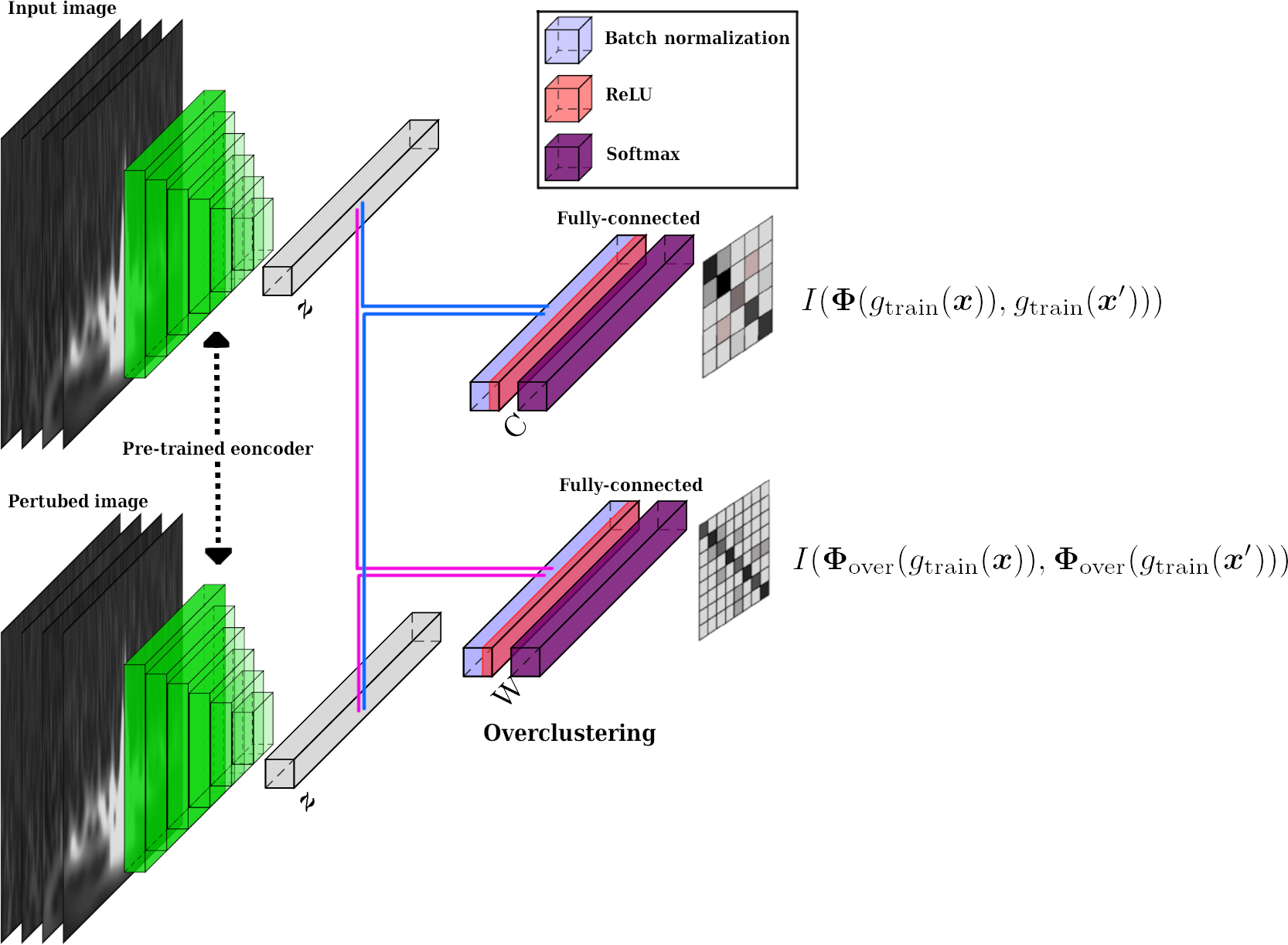}
  \caption{
    Schematic view of the proposed IIC architecture for image classification of transient noise.
    $C \in \mathbb{Z}$ is the estimated number of classes of the transient noise and $W \in \mathbb{Z}$ is the number of classes for overclustering, where $C < W$.
    IIC uses the VAE's pre-trained encoder and classifies transient noise from the latent variables $\z$, and the softmax (also called normalized exponential function) layer outputs the probabilities of classification of transient noise in the range of $[0,1]$.
    The result of the mutual information, including the clustering of $C$ and the overclustering of $W$ are added (at the right end), and then IIC trains to maximize this mutual information.
    The two input images to IIC are the center-cropped image of transient noise and its perturbed image.
  }\label{fig:iic_arch}
\end{figure}

IIC for the classification of transient noise uses VAE's pre-trained encoder as shown in \textbf{\figref{fig:iic_arch}}.
Because the features of transient noise have already been learned by the VAE's pre-trained encoder, the IIC architecture was simply configured with VAE's pre-trained encoder and a fully connected layer.
Additionally, overclustering~\cite{ref:Ji2019} was applied to IIC to improve the performance of the architecture.
Regarding the two inputs $\x, \x^{\prime} \in \mathcal{X}$ to IIC in training, we let $\x$ be the center-cropped 2D image of transient noise and $\x^{\prime}$ be the perturbed 2D image (Section~\ref{sec:outline-of-target-dataset}) in the time direction, whose shapes of images are the same as $(4, 224, 224)$, and $g_{\mathrm{train}}:\mathcal{X} \to \mathcal{Z}$ is the mapping of pre-trained encoder.
Then, IIC trains to maximize the mutual information $I(\bm{\Phi}(g_{\mathrm{train}}(\x)), \bm{\Phi}(g_{\mathrm{train}}(\x^{\prime})))$ and overclustering $I(\bm{\Phi}_{\mathrm{over}}(g_{\mathrm{train}}(\x)), \bm{\Phi}_{\mathrm{over}}(g_{\mathrm{train}}(\x^{\prime})))$, where $\bm{\Phi} \in \mathbb{R}^C$ and $\bm{\Phi}_{\mathrm{over}} \in \mathbb{R}^W$.
We also used several classifiers (typically set as five) and backpropagated the ensemble average for this mutual information to reduce the initial value dependence of the neural networks.
The details are shown in~\secref{sec:training-iic}.

Following IIC training, although we could obtain a classification result from one classifier, the softmax outputs of each classifier differ slightly due to the initial value of the neural networks. Therefore, this result should be averaged by multiple classifiers to obtain a uniform classification result.
Regarding unsupervised learning, the classification labels are given randomly in training.
For example, the first classifier classifies the data as class label ``0,'' while the second classifier classifies the same data as class label ``1.''
Therefore we could not calculate an ensemble average of multiple classification results.
As an alternative, we used an approach that constructs one classification result from the features extracted by multiple classification results using the spectral clustering~\cite{ref:VonLuxburg2007}.
Now, let $D \in \mathbb{Z}$ be the total number of datasets and $K \in \mathbb{Z}$ be the number of classifiers, $C \in \mathbb{Z}$ be the estimated number of classes, and $\mathcal{M}_{D\times C}(\mathbb{R}$)
be the classification matrix output from one classifier.
Based on the \textit{hypermatrix} $\mathcal{H}_{D\times CK}(\mathbb{R})$, which is composed of the concatenated results of multiple classifiers,
we calculated an affinity matrix $\mathcal{A}_{D\times D}(\mathbb{R})$ using the Gaussian similarity function $(A)_{i,j} = \exp{(-||h_i - h_j||^2)}$, where $h_i$ is a row vector of the hypermatrix $\mathcal{H} = (h_1, \dots, h_i, \dots, h_{D})^T$.
After applying spectral clustering to the affinity matrix, we obtained the following new classification matrix: $\mathcal{M}_{D\times C}(\mathbb{R})$.

\section{Training Process and Result} \label{sec:training}
We report on the training process and results of our proposed architecture step by step.  

\subsection{Outline of Dataset} \label{sec:outline-of-target-dataset}
The target dataset is the Gravity Spy dataset of LIGO O1~\cite{ref:Zevin2017}~\cite{ref:Bahaadini2018} which consists of 8535 transient noise images with 22 different labels (e.g., ``Blip,'' ``Power\_Line,'' and ``Koi\_Fish'').
All transient noise in the dataset has been selected with the SNR $\ge 7.5$ by Omicron software~\cite{ref:Robinet2020}.
Each transient noise image is represented as a time-frequency spectrogram 2D image.
One transient noise is recorded in four time durations: 0.5, 1.0, 2.0, and 4.0 s.
More details of the labels, the distribution of the dataset, and 2D images of the transient noise are explained in Figure 1 in ref.~\cite{ref:sakai2021unsupervised}.

We also applied a preprocess that randomly shifts between $\pm$ 0-24 px from the center of the image to the transient noise 2D image.
The 2D images of the transient noise in each time duration have the shape of 224 $\times$ 272 px.
The preprocessing randomly shifts the time direction of the image between 0 and 24 px and crops it at 224 $\times$ 224 px (square image).
The purpose of this preprocessing is to allow the proposed architecture to train transient noise features without relying on small-time shifts.
Reducing the size of input images contributes to saving the training cost.
More details of the preprocessing are shown in Figure 7 in ref.~\cite{ref:sakai2021unsupervised}.

We used these four original and perturbed 2D images as the input data, as shown in Figures~\ref{fig:vae_arch} and~\ref{fig:iic_arch}.
More details of the dataset can be found in ref.~\cite{ref:sakai2021unsupervised}.

\subsection{Training of VAE} \label{sec:training-vae}
The objective of VAE training is to maximize the variational lower bound expressed by \Eqref{eq:variational-lower-bound-formula}.
Regarding the log-likelihood expressed as the second term of \Eqref{eq:variational-lower-bound-formula},
the original study~\cite{ref:Kingma2014} proposed two equations: Bernoulli and Gaussian distributions.
Moreover, we used the mean squared error (MSE) as a reconstruction error.
The three equations are expressed as
\begin{equation}
  \lnp{\bm{\theta}}{\x | \z} =
  \begin{cases}
    \sum^D_i \left[
      x_i \ln y_i + (1-x_i)\ln(1-y_i) 
    \right]\   & \text{(Bernoulli distribution)}\\
    \sum^D_i (x_i - y_i)^2 & \text{(MSE)}\\
    1/2\ln(2\pi) + \sum^D_i [\ln|\sigma_i| + (x_i - \mu_i)^2/(2\sigma^2_i)] & \text{(Gaussian distribution)}
  \end{cases},
  \label{eq:log-likelihood}
\end{equation}
where $y_i$ is each pixel of the reconstructed image, $\mu_i$ and
$\sigma_i$ are the mean and variance with respect to each pixel of $x_i$, respectively.
Because the Gaussian distribution requires both $\mu_i$ and $\sigma_i$, we doubled the output layer of the decoder in \figref{fig:vae_arch}.

Generally, deep learning has hyperparameters for optimal training.
We investigated the hyperparameters as shown in \textbf{\tabref{tab:investigation-table}} (top) and conducted an experiment using our proposed architecture as follows: the initial learning rate (used Adam~\cite{ref:Adam}, which is one of the stochastic gradient descent methods) is in the range of [$5\times 10^{-7}, 5 \times 10^{-2}$] in increments of one digit;
the minibatch size is in the range of [$32, 128$] in increments of $\approx 32$;
the training size is in the range of [60\%, 90\%] in increments of 10\%; 
the dimensions of the latent variable $\z$ are $64, 128, 256, 512,$ and $1024$;
VAE has the evaluation phase every five epochs to quantify the performance of the model, and the size of the evaluation dataset is $(1 -$ Training size), respectively.
In this study, the case of representative hyperparameters, as shown in~\textbf{\tabref{tab:vae-parameters}} was used to explain the training process.

\begin{figure}[t]
  \centering
  \includegraphics[scale=0.60]{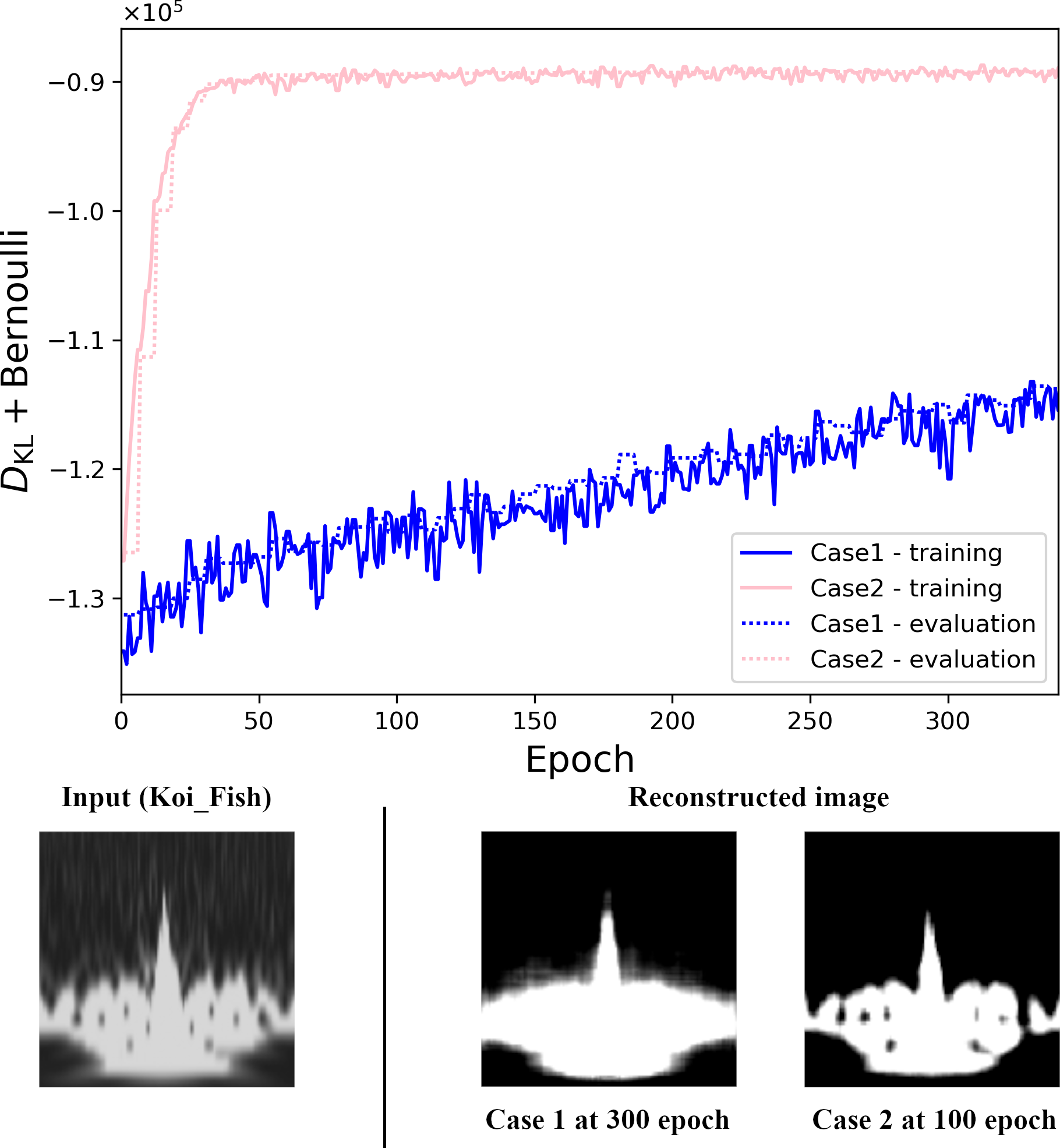}
  \caption{
    VAE training curves using the Bernoulli distribution in \tabref{tab:vae-parameters}  \textbf{(Top)}.
    The decoder's output used the model at the specific epoch \textbf{(Bottom)}.
    An input image with the Gravity Spy label of ``Koi\_Fish'' (bottom left).
    The output image by ``Case 1'' at 300 epochs does not appropriately reconstruct the input image (bottom center).
    The output image by ``Case 2'' at 100 epochs appropriately reconstructs the input image (bottom right).
  }\label{fig:vae-loss-bernoulli}
\end{figure}

The training curves using the Bernoulli distribution in \tabref{tab:vae-parameters} are shown in \textbf{\figref{fig:vae-loss-bernoulli}} (top).
Because ``Case 1'' and ``Case 2'' of evaluation curves (plotted as dashed) are close to these training curves (plotted as solid), there is no overfitting in the training process.
Although the training curve of ``Case 1'' (drawn in blue) progresses slowly, the training curve of ``Case 2'' (drawn in pink) sets a higher learning rate than ``Case 1''.
It also progresses rapidly and is stabilized near 100 epochs.
Considering the Bernoulli case, an optimizer with a small learning rate would be the cause of slow progress in training.
The reconstructed images from the decoders of ``Case 1'' at 300 epochs and ``Case 2'' at 100 epochs are shown in \figref{fig:vae-loss-bernoulli} (bottom).
The ``Case 1'' image may not have been reconstructed appropriately.
However, the ``Case 2'' image seems to extract the feature of the input image appropriately.
Therefore, ``Case 2'' is considered suitable in the case of the Bernoulli distribution.

\begin{figure}[t]
  \centering
  \includegraphics[scale=0.60]{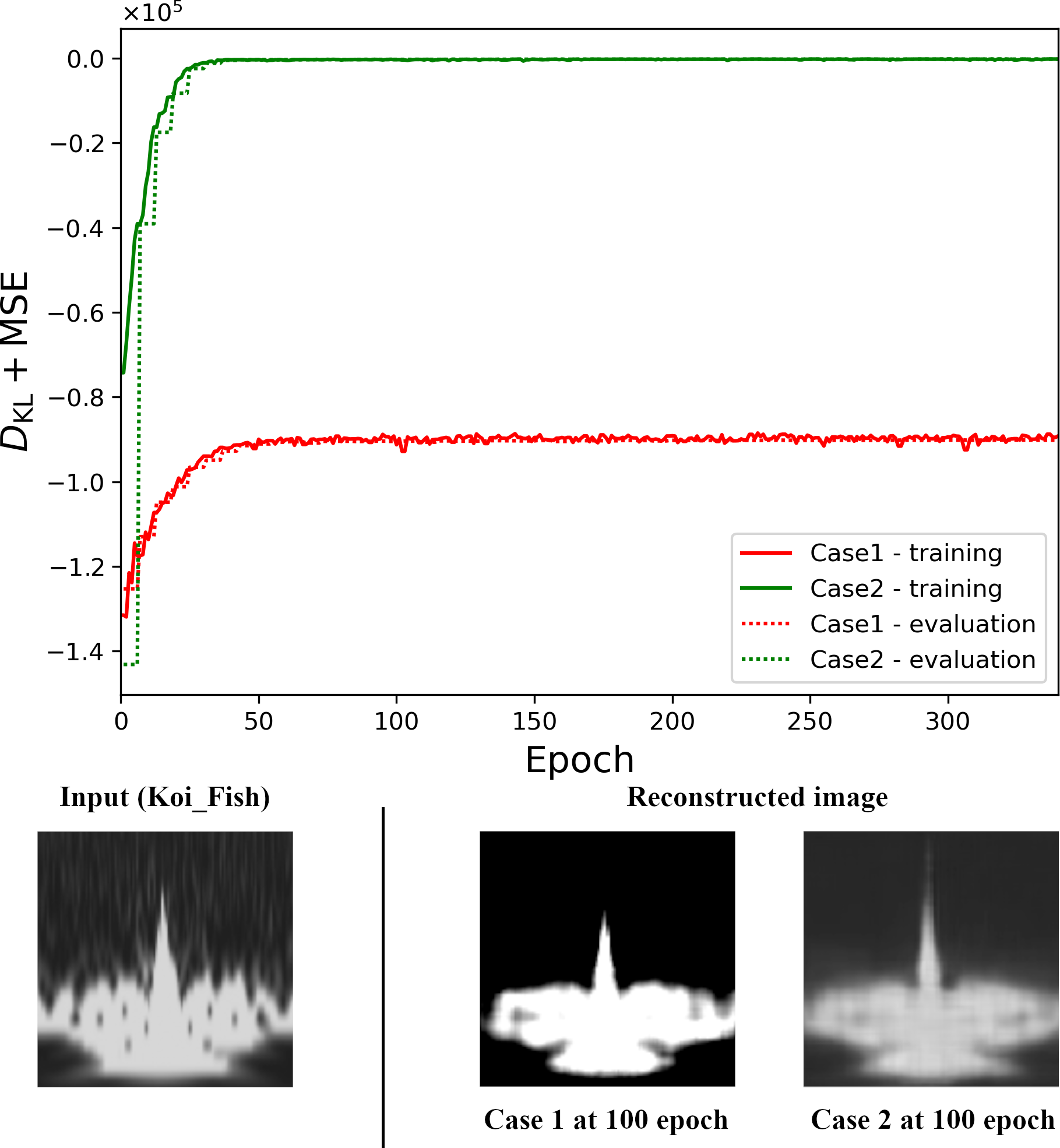}
  \caption{
  Same representation of Figure~\ref{fig:vae-loss-bernoulli} but for MSE.
  The output image by ``Case 1'' at 100 epochs does not appropriately reconstruct the input image (bottom center).
  The output image by ``Case 2'' at 100 epochs appropriately reconstructs the input image (bottom right).
  The training curve of ``Case 1'' is worse than ``Case 2''.
  Regarding ``Case 2'', the training curve is small, but it reconstructs the transient noise along with the background.
  As a result, the reconstructed image seems to be blurred.
  Therefore the MSE architecture is not suitable to reconstruct the transient noise.
  }\label{fig:vae-loss-mse}
\end{figure}

The training curves using MSE in \tabref{tab:vae-parameters} are shown in \textbf{\figref{fig:vae-loss-mse}} (top).
The training curves converge in both cases, with the difference being that the ``Case 2'' curve (drawn in green) is close to zero, and the ``Case 1'' curve (drawn in red) is smaller than the ``Case 2'' curve.
Therefore, we confirmed that ``Case 2'' is more optimal for the hyperparameters than ``Case 1'' in the MSE case.
The reconstructed images of ``Case 1'' at 100 epochs and ``Case 2'' at 100 epochs are shown in \figref{fig:vae-loss-mse} (Bottom).
The ``Case 1'' image would only be roughly reconstructed because it is considered to be caused by insufficient optimization of the hyperparameters.
The ``Case 2'' reconstructed image seems to be close to the input image, but this image is blurry. This is the reason why the decoder of ``Case 2'' reconstructs images along with background features. Therefore, the training curve might be close to zero.

\begin{figure}[t]
  \centering
  \includegraphics[scale=0.60]{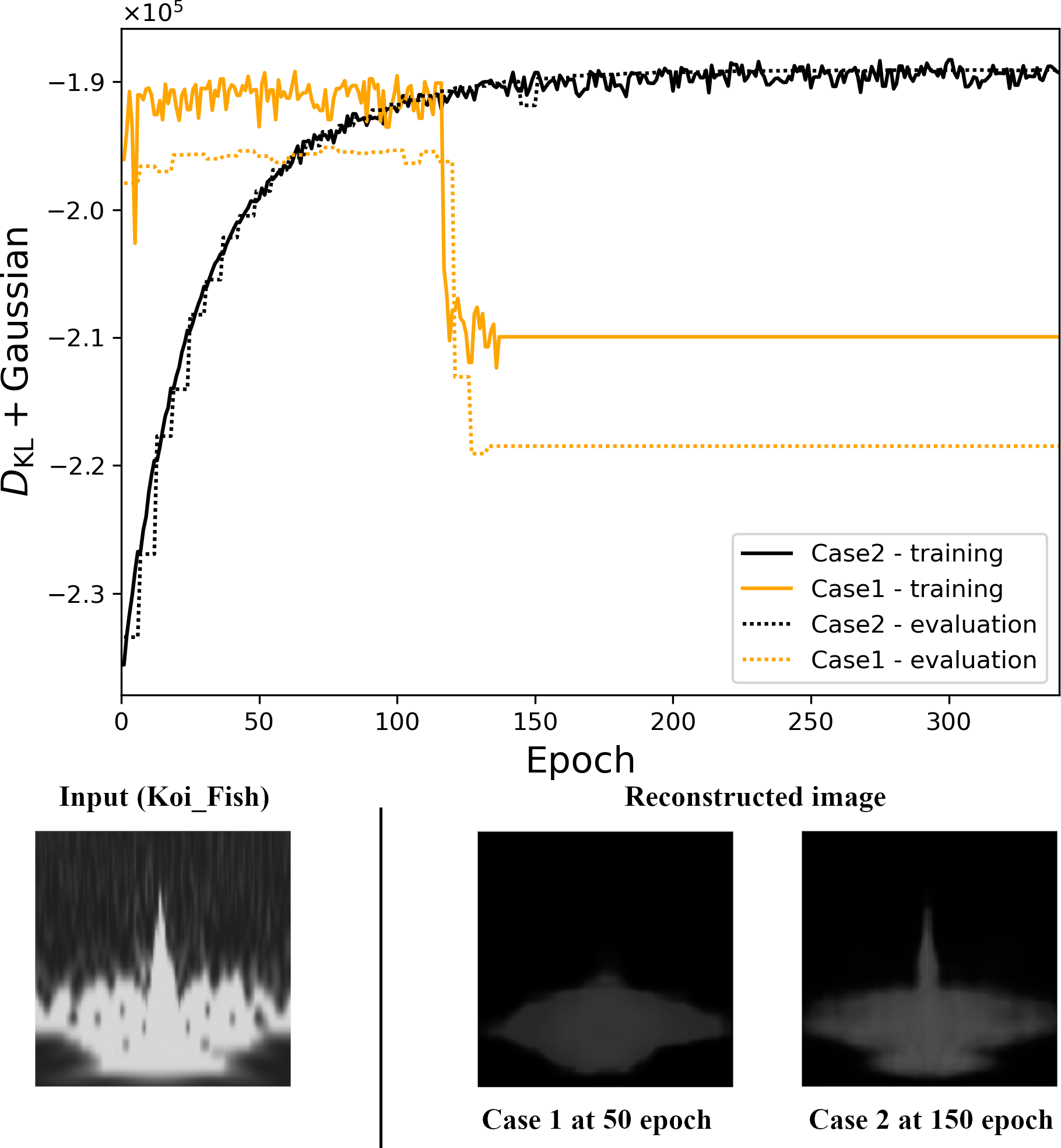}
  \caption{
  Same representation of Figure~\ref{fig:vae-loss-bernoulli} but for the Gaussian distribution.
  An output image by ``Case 1'' at 50 epochs does not appropriately reconstruct the input image (bottom center).
  An output image by ``Case 2'' at 150 epochs approximately reconstructs the input image (bottom right).
  }\label{fig:vae-loss-gaussian}
\end{figure}

The training curves using the Gaussian distribution are shown in \textbf{\figref{fig:vae-loss-gaussian}} (top).
The ``Case 1'' training curve (drawn in yellow) has a difference between the training and evaluation curves, and this training is considered overfitting.
The ``Case 2'' training curve (drawn in black) converges at 150 epochs and the evaluation curve is close to the training curve. ``Case 2'' is stable.
Therefore, ``Case 2'' is more appropriate than ``Case 1'' in the Gaussian case.
The reconstructed images ``Case 1'' at 50 epochs and ``Case 2'' at 150 epochs, which are $\mu_i$ of the decoder output in \Eqref{eq:log-likelihood}, is shown in \figref{fig:vae-loss-gaussian} (Bottom).
Both images would be less clear than the input image.
The reconstructed image is averaged because the mean $\mu_i$ of the Gaussian distribution was applied to the output.

We used two NVIDIA GeForce RTX 2080 Ti GPUs, Intel Xeon CPU E5-2637 v4 (eight cores), and with 125 GB of main memory. The computational cost of each training process is shown in \tabref{tab:vae-parameters}.
The computational cost increases as the batch size and training size increase.
Bernoulli and MSE cases have almost the same computational cost because the neural network layers are the same.
However, the computational cost for the Gaussian case is higher than that for the Bernoulli and MSE cases.
The reason for the higher computational cost in the Gaussian case is that the decoder optimizes the neural network through the mean and variance of the output, which increases the neural network layers.

\begin{table}[t]
\caption{
    Hyperparameters on training of VAE (top) and IIC (bottom).
}\label{tab:investigation-table}
\centering
\begin{tabular}{llcc}
\hline
                                     & Hyperparameter name                    & Range                                & Increment value  \\ \hline\hline
\multicolumn{1}{l|}{Training of VAE} & Initial learning rate                  & $[5\times 10^{-7}, 5\times 10^{-2}]$ & $10^{-1}$          \\
\multicolumn{1}{l|}{}                & Mini-batch size                        & $[32, 128]$                          & $\approx 32$ \\
\multicolumn{1}{l|}{}                & Training size                          & $[60\%, 90\%]$                       & $10\%$             \\
\multicolumn{1}{l|}{}                & Dimensions of $\z$ & $64, 128, 256, 512, 1024$            & -                  \\ \hline\hline
\multicolumn{1}{l|}{Training of IIC} & Initial learning rate                  & $[5\times 10^{-7}, 5\times 10^{-2}]$ & $10^{-1}$          \\
\multicolumn{1}{l|}{}                & Mini-batch size                        & $[64, 256]$                          & $32$               \\
\multicolumn{1}{l|}{}                & Number of classes                      & $[22, 100]$                          & $2$                \\
\multicolumn{1}{l|}{}                & Number of over clustering              & $[50, 500]$                          & $50$               \\
\multicolumn{1}{l|}{}                & Classifier number                      & $3, 5, 10, 20$                       & -                  \\ \hline
\end{tabular}
\end{table}
\begin{table}[t]
\caption{
    Representative hyperparameters for the training process of the proposed architecture and its computational cost.
    All cases are adopted by Adam optimization~\cite{ref:Adam}.
}\label{tab:vae-parameters}
\centering
\begin{tabular}{llccccc}
\hline
                                                &       & \multicolumn{1}{l}{Initial learning rate} & \multicolumn{1}{l}{Batch size} & \multicolumn{1}{l}{Training size} & \multicolumn{1}{l}{Dimension of $\z$} &  \multicolumn{1}{l}{Computational cost}\\ 
                                                & & & & & & (500 epochs)\\
\hline \hline
\multicolumn{1}{c}{\multirow{2}{*}{Bernoullie}} & Case1 & $5 \times 10^{-5}$                               & 128                            & 80\%                              & 512                                &5.0 h\\
\multicolumn{1}{c}{}                            & Case2 & $5 \times 10^{-4}$                               & 128                            & 80\%                              & 512                                &5.0 h\\ \hline
\multirow{2}{*}{MSE}                            & Case1 & $5 \times 10^{-5}$                               & 64                             & 60\%                              & 256                                &2.0 h\\ 
                                                & Case2 & $5 \times 10^{-4}$                               & 96                             & 70\%                              & 512                                &3.5 h \\ \hline
\multirow{2}{*}{Gaussian}                       & Case1 & $5 \times 10^{-4}$                               & 32                            & 60\%                              & 128                               &  1.8 h\\
                                                & Case2 & $5 \times 10^{-2}$                               & 82                             & 80\%                              & 256                                &  6.0 h\\ \hline
\end{tabular}
\end{table}

We also investigated the latent variables, which are outputs from the encoder.
The auxiliary analysis is frequently applied to a training model to understand the learned features of the dataset.
Latent variables are important for the representation of the dataset. However, because the latent space is generally the higher dimension, it is difficult to assess its utility.
For this reason, visualization methods are frequently applied by embedding from a high-dimensional space to a lower-dimensional space.
A typical approach is principal component analysis (PCA), which is a linear dimension reduction technique based on keeping the covariance of the data.
Alternatively, the t-distributed stochastic neighbor embedding (t-SNE)~\cite{ref:Roweis2000} is a nonlinear technique that preserves the distance between the higher-dimensional and lower-dimensional features.
PCA leads to a better understanding of the data’s global structure. Alternately, t-SNE helps in the understanding of the local structure which is important for comprehending clustering visually.
Therefore, we applied the t-SNE method for the visualization of the latent space.
Let $p_{ij} \to \mathbb{R}$ be the joint distribution of the two latent variables $\z_i, \z_j \in \mathcal{Z}$ and $q_{ij} \to \mathbb{R}$ be the joint probability of the lower -dimensional variables $\y_i, \y_j \in \mathcal{Y}$, where $\mathcal{Y}$ is the lower -dimensional space.
Suppose $p_{ij}$ follows the Gaussian distribution and $q_{ij}$ follows t-distribution, respectively, then, the t-SNE objective $d$ is determined using the following equation
\begin{equation}
  d = \sum_{ij}p_{ij}\log \frac{p_{ij}}{q_{ij}}.
\end{equation}

\begin{figure}[t]
  \centering
  \includegraphics[scale=1.3]{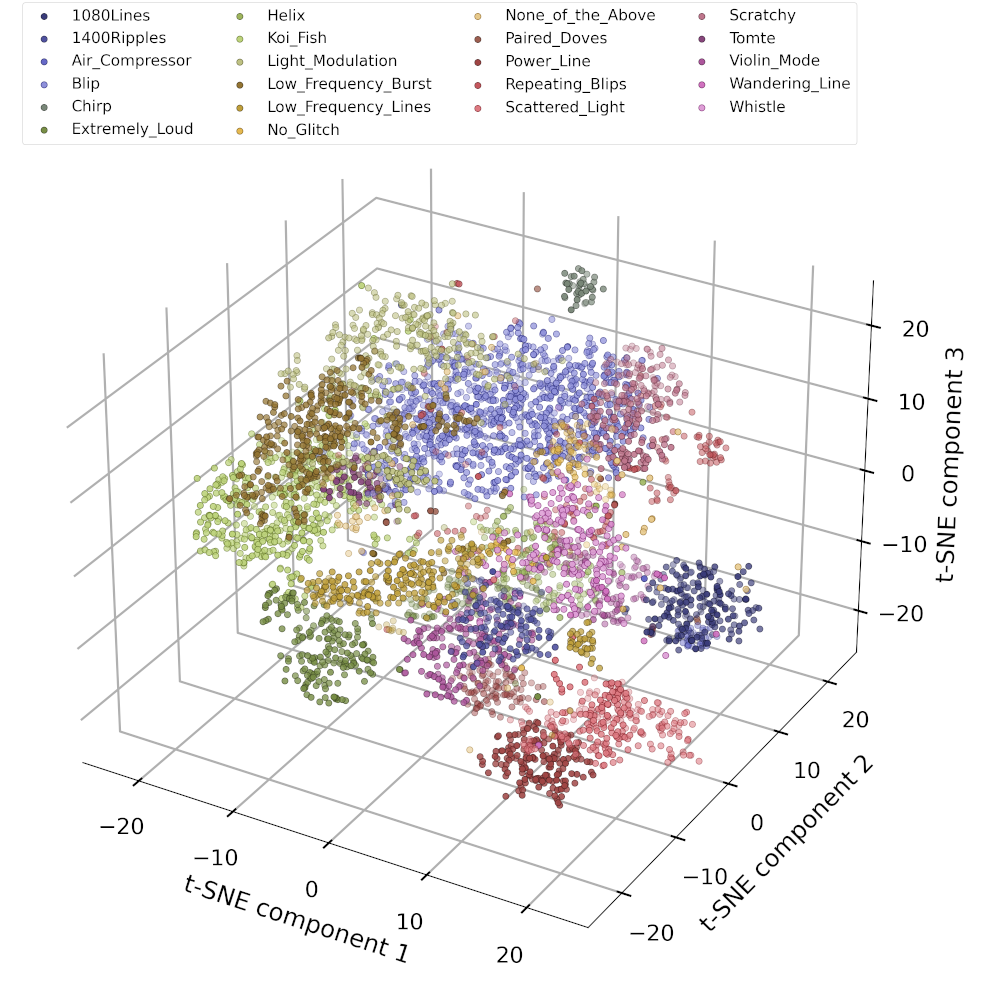}
  \includegraphics[scale=0.25]{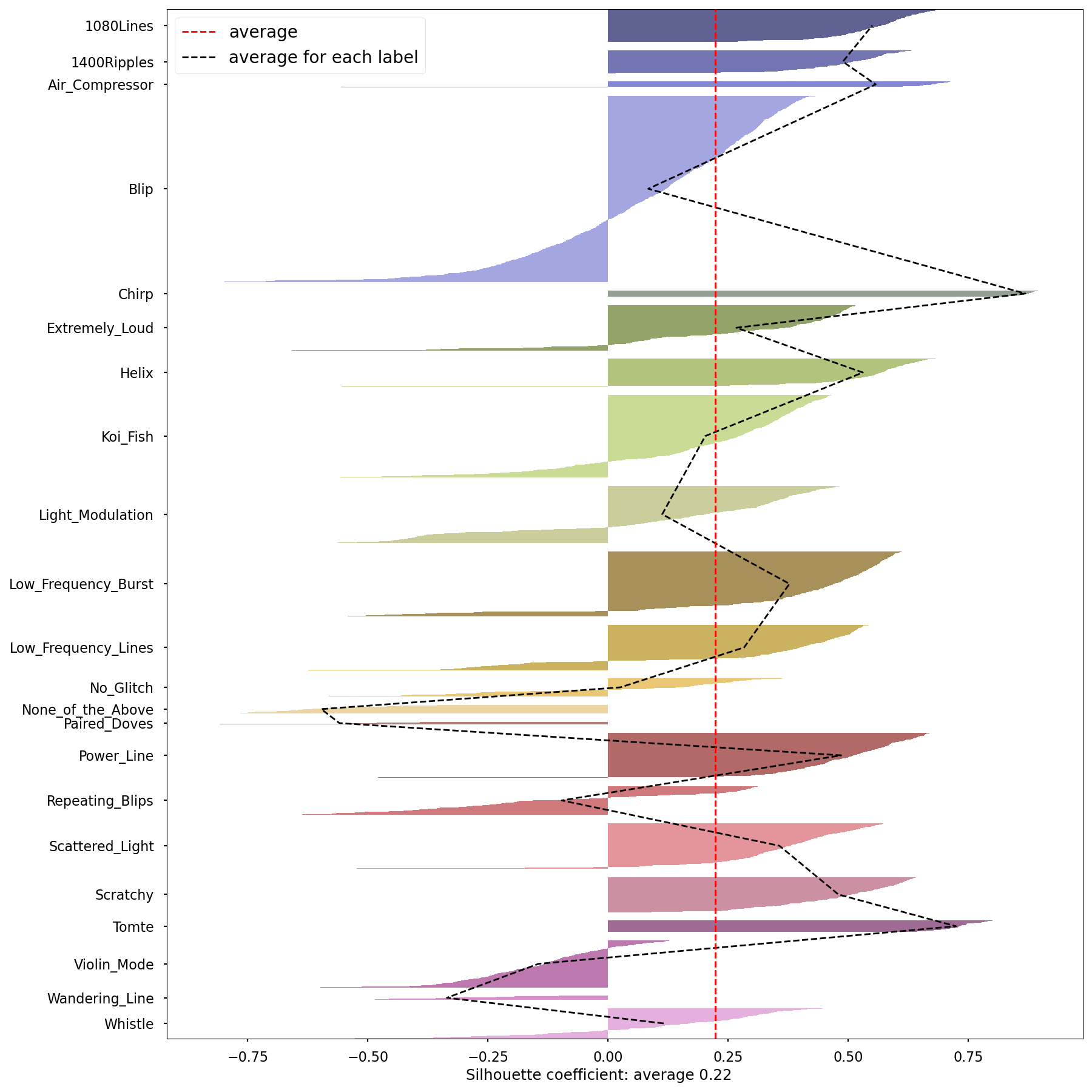}
  \caption{
  Colors represents 22 types of Gravity Spy label.
  The t-SNE mapping with Bernoulli ``Case 2'' at 100 epochs \textbf{(Top)}.
  The silhouette coefficient of the above t-SNE results \textbf{(Bottom)}. 
  A dashed line drawn in black represents the average silhouette coefficient of each label, and the dashed red line represents an average for all labels.
  }\label{fig:bern-tsne-silhoutte}
\end{figure}

\begin{figure}[t]
  \centering
  \includegraphics[scale=1.3]{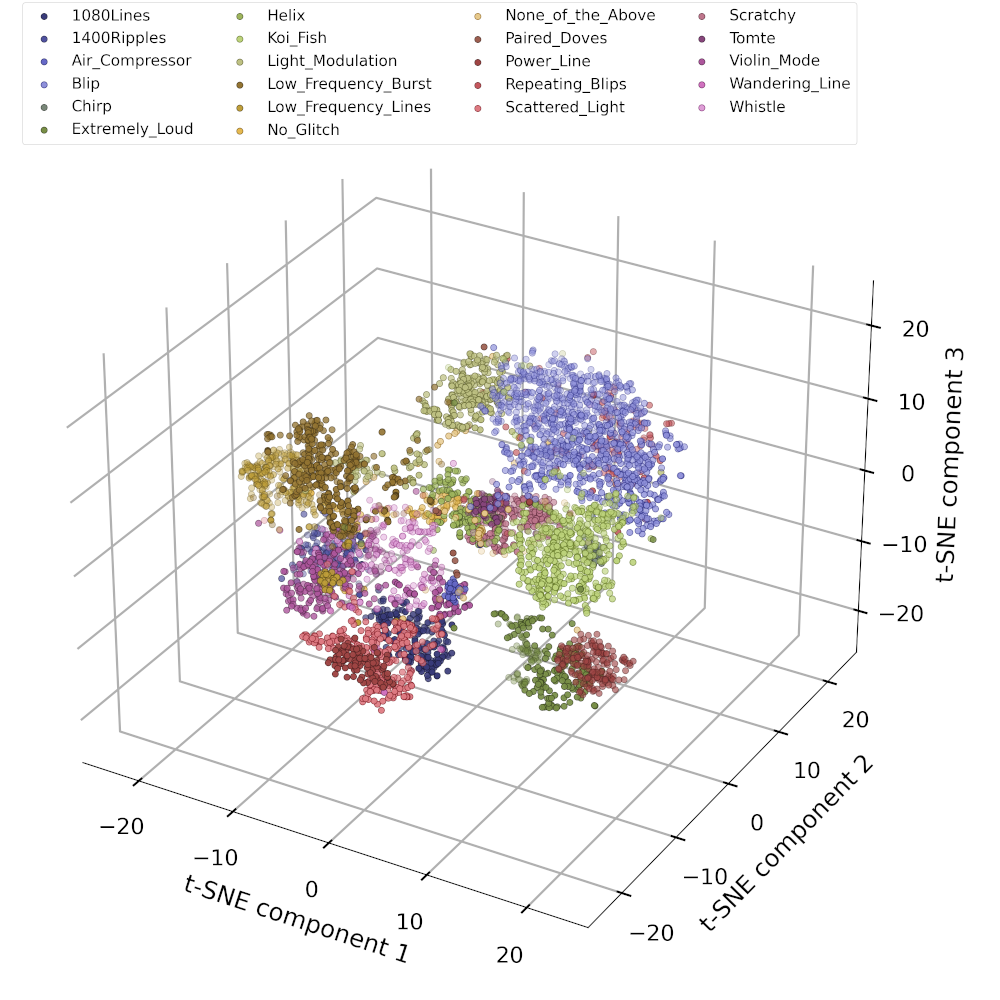}
  \includegraphics[scale=0.25]{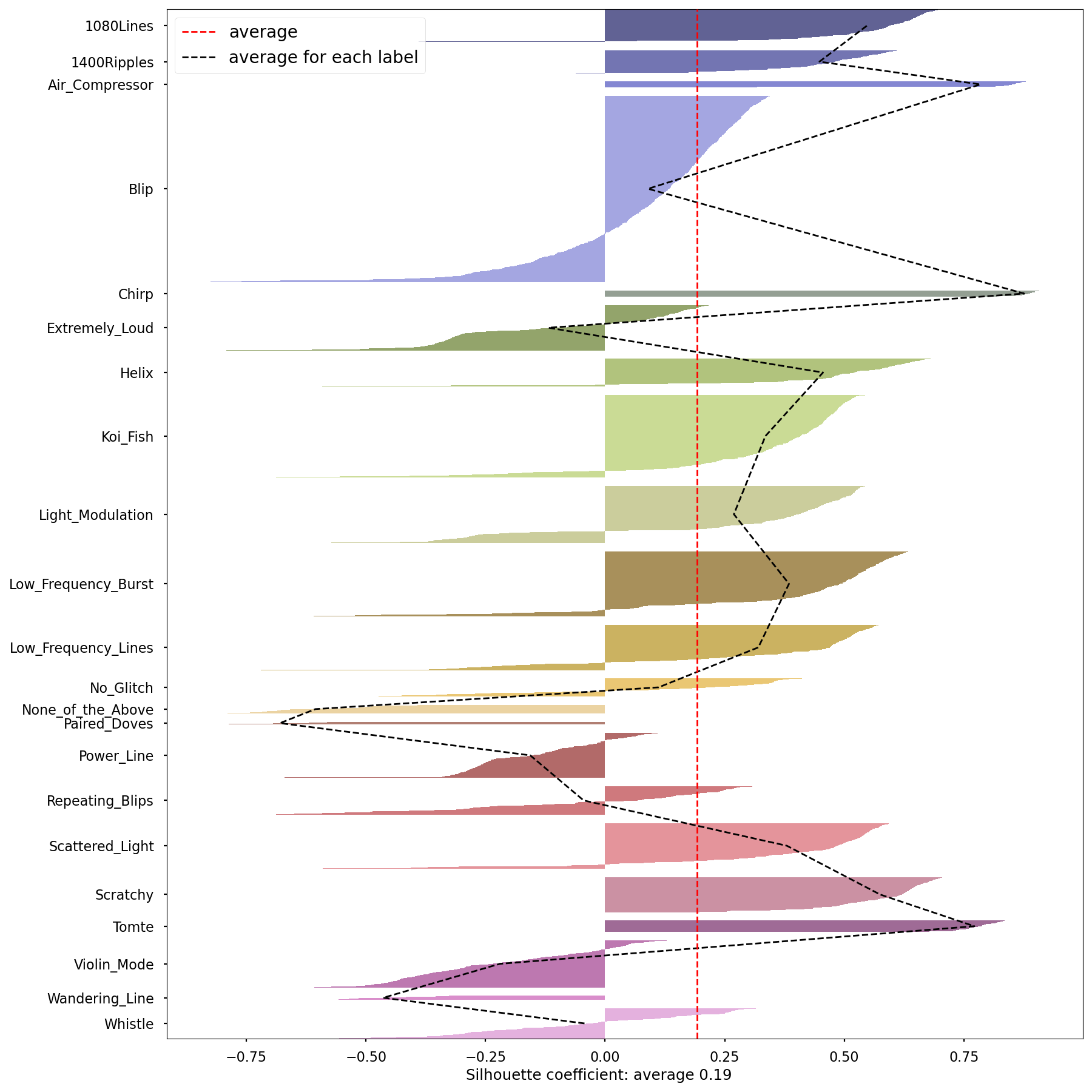}
  \caption{
  Same representation of Figure~\ref{fig:bern-tsne-silhoutte} but for the MSE ``Case 2'' at 100 epochs.
  }\label{fig:mse-tsne-silhoutte}
\end{figure}

\begin{figure}[t]
  \centering
  \includegraphics[scale=1.3]{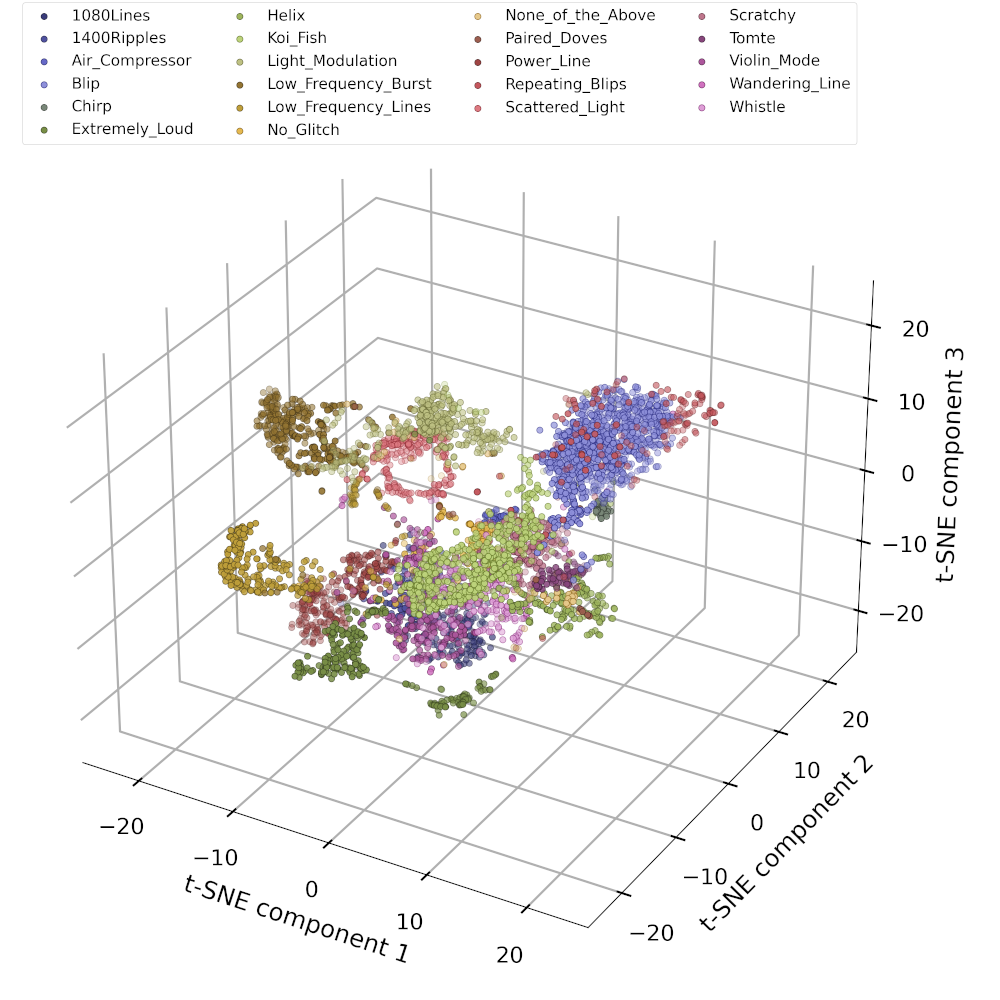}
  \includegraphics[scale=0.25]{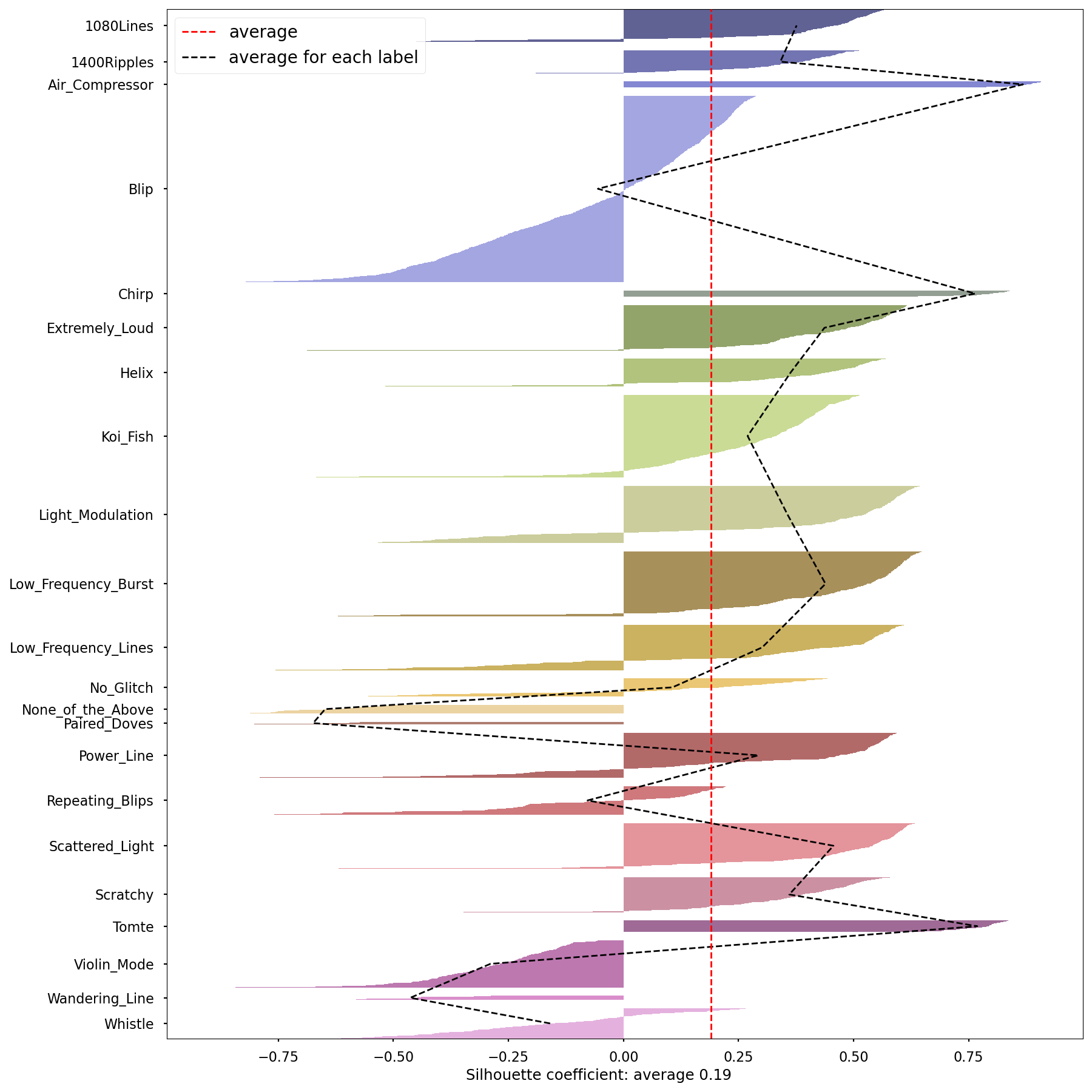}
  \caption{
   Same representation of Figure~\ref{fig:bern-tsne-silhoutte} but for the Gaussian ``Case 2'' at 150 epochs.
  }\label{fig:gaussian-tsne-silhoutte}
\end{figure}

Let $\mathcal{Y} \subset \mathbb{R}^3$ be the lower-dimensional space, and the learning parameters of t-SNE be set as below: perplexity, which is related to the number of nearest neighbors as $40$, the maximum number of iterations for the optimization as $2000$, and random seed as $10$.
We also used the Gravity Spy labels to visualize how the Gravity Spy dataset is clustered in the latent space.
The t-SNE mapping was applied to each model: Bernoulli ``Case 2'' at 100 epochs, MSE ``Case 2'' at 100 epochs, and Gaussian ``Case 2'' at 150 epochs.
The t-SNE results of the Bernoulli, MSE, and Gaussian cases are shown in \textbf{\figref{fig:bern-tsne-silhoutte}} (top) , \textbf{\figref{fig:mse-tsne-silhoutte}} (top), and \textbf{\figref{fig:gaussian-tsne-silhoutte}} (top), respectively.
Considering the Bernoulli case, the data in each class appear to be clustered in 3D space, and each cluster has a spatially expansive structure.
However, considering the MSE case, the data are also clustered, and the difference is that the structure of each cluster seems to be smaller than that of the Bernoulli case.
Similar aspects to the MSE were found in the Gaussian case.

Moreover, the t-SNE results are quantitatively examined using the silhouette coefficient.
Let $\mathcal{C}_{\mathrm{in}} \subset \mathcal{Y}$ be the intra-cluster dataset and $\mathcal{C}_{\mathrm{near}} \subset \mathcal{Y}$ be the nearest-cluster dataset, where $\mathcal{C}_{\mathrm{in}} \cap \mathcal{C}_{\mathrm{near}} = \emptyset$.
The mean intra-cluster distance of the $i$th data is defined as follows: $a^{(i)} = 1/(|\mathcal{C}_{\mathrm{in}}| -1) \sum_{\bm{y}^{(j)}\in \mathcal{C}_{\mathrm{in}}} || \bm{y}^{(i)} - \bm{y}^{(j)}||$, and the mean nearest-cluster distance of the $i$th data is defined as follows: $b^{(i)} = 1/|\mathcal{C}_{\mathrm{near}}| \sum_{\bm{y}^{(j)}\in \mathcal{C}_{\mathrm{near}}} || \bm{y}^{(i)} - \bm{y}^{(j)}||$.
The silhouette coefficient $s^{(i)}$ is defined as
\begin{equation}
  s^{(i)} = \frac{b^{(i)} - a^{(i)}}{\max{(b^{(i)}, a^{(i)})}}.
\end{equation}
Note that the silhouette coefficient is a mapping $\mathcal{Y} \to [-1,1]$ and it approaches $1$ when the clustering is condensing and approaches $0$ when the number of clusters is not appropriate.
The negative value means that these data overlapped with each cluster.

The silhouette coefficients for the Bernoulli, MSE, and Gaussian cases are shown in \figref{fig:bern-tsne-silhoutte} (bottom), \figref{fig:mse-tsne-silhoutte} (bottom), and \figref{fig:gaussian-tsne-silhoutte} (bottom), respectively.
The average values (drawn in red) for all classes are $0.22$ in the Bernoulli case, $0.19$ in the MSE case, and $0.19$ in the Gaussian case. The average silhouette coefficient in the Bernoulli case is greater than that in the other two cases.
The MSE and Gaussian cases overlapped with other classes because their clustering was more condensed than the Bernoulli case. As a result, their silhouette coefficients decreased.

We summarize the results of VAE’s training process. 
The training optimization was conducted in \Eqref{eq:log-likelihood} using three types of the reconstruction error.
The Bernoulli ``Case 2'' decoder at 100 epochs was used to reconstruct the features of transient noise. 
We also confirmed the t-SNE results and silhouette coefficients. As the results show, it was also suggested that the Bernoulli ``Case 2'' encoder at 100 epochs was better for transient noise classification. Therefore, we used the Bernoulli ``Case 2'' encoder at 100 epochs as a pre-trained encoder for IIC.

\subsection{Training of IIC} \label{sec:training-iic}
Following VAE training, IIC uses a pre-trained encoder to classify the transient noise 2D image.
The objective of IIC training is to maximize the mutual information expressed by \Eqref{eq:ojective_mutual_information}.
The maximum mutual information expressed by \Eqref{eq:maximum_mutual_information} increases monotonically with the number of classes.
We divided \Eqref{eq:ojective_mutual_information} by \Eqref{eq:maximum_mutual_information}, and obtained the normalized mutual information $I(\bm\Phi(\x), \bm\Phi(\x^{\prime})) / \ln C$ for the number of classes $C$.
Our proposed architecture maximizes the normalized mutual information in the training process.
This normalized expression allows for mutual information comparisons with the different numbers of classes.

The hyperparameters for IIC are also investigated as shown in \tabref{tab:investigation-table} (bottom) and experimented using our proposed architecture as follows:
the initial learning rate is in the range of [$5\times 10^{-7}, 5 \times 10^{-2}$] in increments of one digit (same as VAE training);
the minibatch size is in the range of [$64, 256$] in increments of $32$;
the number of classes is in the range of [$22, 100$] in increments of $2$;
the number of over clustering is in the range of [$50, 500$] in increments of $50$;
the classifier number is one of $3,5,10,$ and $20$.

\begin{figure}[t]
  \centering
  \includegraphics[scale=0.70]{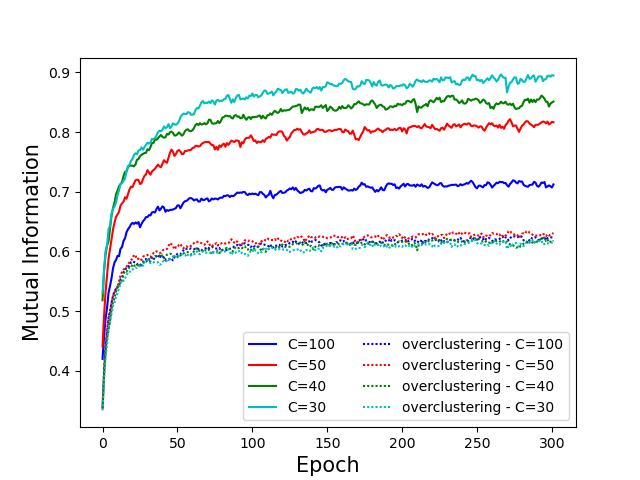}
  \caption{
  Training curves of IIC.
  The solid lines represent the average mutual information for each number of classes $C$.
  The dashed lines represent the average mutual information with overclustering corresponding to the number of classes $C$, and the classes of all overclustering are set to $W=250$.
  }\label{fig:iic-loss}
\end{figure}

Regarding the initial learning rate, when other hyperparameters are fixed, the normalized mutual information is not affected unless the value is far away, for example, $5\times 10^{-7}$ and $5 \times 10^{-2}$.
Similar aspects to the number of overclustering, the classifier number, and the minibatch size were found in the training process.
Therefore, in this study, we set the initial learning rate to $5\times 10^{-4}$, the number of classifiers to five, the number of over clustering to $250$, and the minibatch size to $128$.
We investigated the range of [$22, 100$] to determine the number of classes that affect the normalized mutual information.
Because the previous studies~\cite{ref:Soni2021}~\cite{ref:george2018}~\cite{ref:bahaadini2018direct} presented new classes in addition to the 22 classes of the same dataset, the representative training curves for $C=30, 40, 50,$ and $100$ are shown in \textbf{\figref{fig:iic-loss}}.
Note that because the training curve was oscillating and covered by other training curves, we applied the exponential moving average to these training curves, and the average span was set to $10$ epochs in \figref{fig:iic-loss}.
All the cases converge and stabilize near 100 epochs, as shown in \figref{fig:iic-loss}.
The training curves of over clustering were found to be independent of the number of classes $C$, and the normalized mutual information was high at $C=30$ and $C=40$.
Therefore, from \figref{fig:iic-loss}, the number of classes $C$ is considered to be suitable between $30$ and $40$ in our proposed architecture.

\subsection{Correspondence to Result of Supervised Learning (Gravity Spy labels) and Summary of Obtained Result in Ref.~\cite{ref:sakai2021unsupervised}}
\begin{figure}[t]
  \centering
  \includegraphics[scale=0.70]{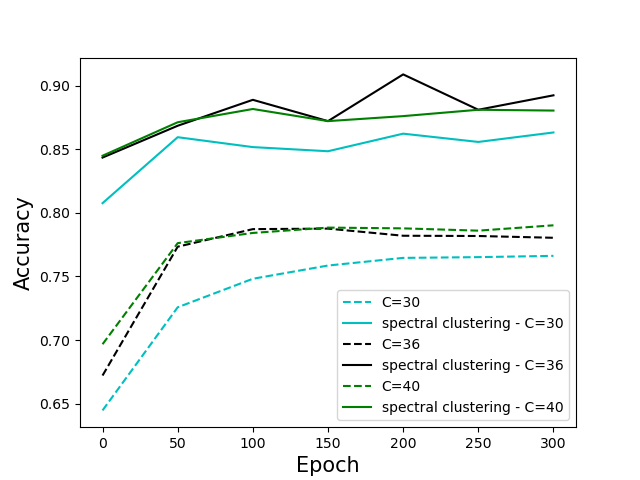}
  \caption{
  Accuracy of the proposed architecture (unsupervised learning) using the Gravity Spy label (supervised learning).
  The colored solid lines show the results of the cases where the number of classes is $30$, $36$, and $40$ (denoted by $C$ in the legend) with the spectral clustering of five classifiers.
  The dashed line represents the average accuracy of five classifiers.
  }\label{fig:iic-accuracy-sc}
\end{figure}

After the IIC training discussed in \secref{sec:training-iic}, we investigated the correspondence between the results of supervised learning (Gravity Spy labels) and our unsupervised learning. Although the details can be found in ref.~\cite{ref:sakai2021unsupervised}, we briefly review the results.
We used the accuracy in Equation~(1) of ref.~\cite{ref:sakai2021unsupervised} defined as follows, $\sum_{i}^C \max(\bm{v}^{(i)})/|\sum_j^{C^\prime}{v}^{(i)}|$, where $C$ is the number of classes with unsupervised learning, $C^{\prime}=22$ is the number of classes with supervised learning (Gravity Spy labels), and $\bm{v}^{(i)}$ is the $i$th column in the confusion matrix (e.g., in Figure~4 of ref.~\cite{ref:sakai2021unsupervised}).
By calculating the accuracy as a criterion, we determined the number of classes in unsupervised learning.
Furthermore, we used the spectral clustering discussed in \secref{sec:architecture} to compress the multiple results of classification into one result. 
We investigated the number of classes between $30$ and $40$, which yields high values for the normalized mutual information in \secref{sec:training-iic}.
The results for the representative numbers $C=30, 36,$ and $40$ are shown in \textbf{\figref{fig:iic-accuracy-sc}}.
The dashed line represents the average accuracy from each classifier, which was set to five classifiers, and the solid line represents the accuracy using the spectral clustering for the results of five classifiers.
In all cases, spectral clustering outperformed the average of the five classifiers, with the highest accuracy of $90.9\%$  achieved with the number of classes $C=36$ at $200$ epochs.
Note that the Gravity Spy project~\cite{ref:Zevin2017} using supervised learning achieved $97.1\%$ accuracy on the testing data using the same dataset used here.
In terms of accuracy, the proposed architecture can classify with performance close to supervised learning, and also has the advantage that unsupervised learning does not require the data annotations.
The proposed architecture shows the results of supervised learning with a high level of consistency for classification.
For example, the data of ``1080Lines'' were classified into one class for both supervised and unsupervised learning.
Furthermore, our architecture indicates the existence of the unrevealed classes from the Gravity Spy dataset.
For example, the ``Blip'' data that were classified as one class in supervised learning would be separated into more detailed subclasses in unsupervised learning.

We confirmed the consistency between the label annotated by the Gravity Spy project and the class provided by our proposed unsupervised-learning architecture and provided the potential for the existence of the unrevealed classes.
More details on the result and discussion of unsupervised classification are shown in ref.~\cite{ref:sakai2021unsupervised}.

\section{Summary} \label{sec:summary}
Transient noise appearing in the data from gravitational-wave detectors frequently causes problems. Because transient noise is considered to be associated with the environment and instrument, classifying it may help us understand its origin and improve the detector's performance. In our previous study~\cite{ref:sakai2021unsupervised}, an architecture of classifying the transient noise using a time-frequency 2D image (spectrogram), which uses the unsupervised deep learning combined with VAE and IIC was proposed.
The training process of unsupervised learning is essential. In this study, using the Gravity Spy dataset, we reported on how the training process of unsupervised-learning architecture of ref.~\cite{ref:sakai2021unsupervised} was conducted.
We used three types of objective functions in the training process for feature learning of transient noise using VAE.
We investigated the three types of training curves, reconstructed images from the decoders, t-SNE mapping of the latent variables, and silhouette coefficient of t-SNE mapping.
The normalized objective function was used in the training process using IIC for transient noise classification, and the results were compared between the different classes to investigate the appropriate number of classes.
We also confirmed that the classification result compressed from the multiple unsupervised classifications approaches the accuracy in supervised learning.

A more detailed discussion of the results for the classification of transient noise using our proposed unsupervised architecture can be found in ref.~\cite{ref:sakai2021unsupervised}.

We applied the unsupervised classification to the Gravity Spy dataset of LIGO O1 as a first step. In future work, we will apply our architecture to the recent observation run (O2 and O3) dataset.
Furthermore, we will use our unsupervised classification in KAGRA to develop a transient noise system.
Additionally, we will extend our architecture to self-supervised learning~\cite{ref:lee2013pseudo} to improve classification accuracy, in which the architecture generates \textit{pseudo labels} for a given dataset and re-trains it.

\medskip

\medskip
\vspace{1cm}
\textbf{Acknowledgements} \par 
The authors grateful to the members of the Gravity Spy project for enlightening discussions. 
This study was supported in part by the Inter-University Research Program of the Institute for Cosmic Ray Research, University of Tokyo, Japan. 
It was also supported in part by the Japan Society for the Promotion of Science (JSPS) Grants-in-Aid for Scientific Research on Innovative Areas, Grant No.~24103005 [JP17H06358, JP17H06361, and JP20H04731], by JSPS Core-to-Core Program A, Advanced Research Networks, and by JSPS KAKENHI [Grant No.~19H0190 (Y.I. and H.T.) and Nos.~19K14636 and 21H05599 (Y.~Shikano.)],  by JST, PRESTO [Grant No. JPMJPR20M4 (Y.~Shikano.) ].


\medskip
\textbf{Conflict of Interest} \par
The authors declare no conflict of interest.

\medskip

%

\bibliographystyle{MSP}
\bibliography{references}

\newpage


\end{document}